\documentclass{svproc}

\usepackage{graphicx}

\usepackage{amssymb}

\usepackage{amsmath,braket,placeins}
\usepackage[sort&compress]{natbib}
\usepackage[colorlinks=true,citecolor=blue,linkcolor=red,urlcolor=red]{hyperref}

\newcommand{\bq}{\begin{equation}} \newcommand{\eq}{\end{equation}}
\newcommand{\bqali}{\bq\begin{aligned}}
\newcommand{\eqali}{\end{aligned}\eq}
\newcommand{\bqn}{\begin{equation*}}
\newcommand{\eqn}{\end{equation*}}

\newcommand\D{\operatorname{d}\!}

\newcommand\z{{\bf z}}

\newcommand\x{{\bf x}}
\newcommand\y{{\bf y}}

\newcommand\kb{k_\text{\tiny B}}

\newcommand\rC{r_\text{\tiny C}}

\graphicspath{{../figures/}}

\begin{document}

\title{Collapse models: main properties and the state of art of the experimental tests}
\titlerunning{Collapse models, properties and experimental tests}  %  (for running head)

\author{Matteo Carlesso\inst{1} \and Sandro Donadi\inst{2}}
\authorrunning{Matteo Carlesso et al.} % abbreviated author list (for running head)
\tocauthor{Matteo Carlesso, and Sandro Donadi}
\institute{Department of Physics, University of Trieste,\\
Strada Costiera 11, 34151 Trieste, Italy\\
Istituto Nazionale di Fisica Nucleare, Via Valerio 2, 34127 Trieste, Italy\\
          \email{matteo.carlesso@ts.infn.it}
\and
Frankfurt Institute for Advanced Studies (FIAS),\\
Ruth-Moufang-Stra$\ss$e 1, 60438 Frankfurt am Main, Germany}
\maketitle

\date{\today}
\begin{abstract}
{Collapse models represent one of the possible solutions to the measurement problem. These models modify the Schr\"{o}dinger dynamics with non-linear and stochastic terms, which guarantee the localization in space of the wave function avoiding macroscopic superpositions, like that described in the Schr\"{o}dinger's cat paradox. The Ghirardi-Rimini-Weber (GRW) and the Continuous Spontaneous Localization (CSL) models are the most studied among the collapse models. Here, we briefly summarize the main features of these models and the advances in their experimental investigation.}
\end{abstract}
%
%\pacs{}

\section{Introduction}

Quantum Mechanics is the most precise theory we have for describing the microscopic world. However, since its formulation, the theory never stopped to raise issues regarding its meaning. In particular, the superposition principle does not seem to apply to the macroscopic world. This raises the well-known measurement problem.\\

Collapse models provide a phenomenological solution to such a problem. These models modify the Schr\"{o}dinger equation by adding stochastic and non-linear terms, which implement the collapse of the wave function~\cite{Bassi:2003aa}. An in-built amplification mechanism ensures that their action is negligible for microscopic systems, and become stronger when their mass increases thus providing a natural implementation of the quantum-to-classical transition.\\

The most supported among collapse models are the Ghirardi-Rimini-Weber (GRW) \cite{Ghirardi:1986aa} and the Continuous Spontaneous Localization (CSL) models \cite{Pearle:1989aa, Ghirardi:1990aa}. Their action is determined by two parameters: the collapse rate $\lambda$, and the correlation length of the noise $\rC$. Different theoretical proposals for their numerical value were suggested: $\lambda=10^{-16}\,$s$^{-1}$ and $\rC=10^{-7}\,$m by Ghirardi, Rimini and Weber \cite{Ghirardi:1986aa}; $\lambda=10^{-8\pm2}\,$s$^{-1}$ for $\rC=10^{-7}\,$m, and $\lambda=10^{-6\pm2}\,$s$^{-1}$ for $\rC=10^{-6}\,$m by Adler \cite{Adler:2007ab}. Since these models are phenomenological, the value of their parameters can be bounded and eventually identified only by experiments. \\

The paper is organized as follows: in section \ref{sec:GRW}, we review the GRW model and discuss its main features and how the model provides a solution to the measurement problem. In section \ref{sec:CSL}, we introduce the CSL model and analyze its properties. In sections \ref{int} and \ref{non_int} we briefly review the current experimental attempts to determine the values of the parameters $\lambda$ and $\rC$.  In section \ref{gen} we discuss the dissipative and non-Markovian generalizations of these models. Finally, in section \ref{prop}, we discuss new proposals to set new bounds on these models.

\section{The GRW model}\label{sec:GRW}
The Ghirardi-Rimini-Weber (GRW) model represents the first consistent model where the dynamics induces spontaneous collapses in space. In the GRW model, the wave function of any system is subject to random and spontaneous localizations in space. These collapses are designed in such a way that one recovers the Born rule. Due to an in-built amplification mechanism, the rate of collapses increases with the size of the systems. This guarantees that macroscopic objects always have well-defined positions. Conversely to other collapse models, as for the CSL model (cf.~section \ref{sec:CSL}), the GRW model is not formulated using stochastic differential equations\footnote{It possible to define the model also through a stochastic differential equation describing the interaction with a Poissonian noise, see \cite{Smirne:2014aa,Toros:2016aa}.}, making it ideal to intuitively explain the main features of collapse models. \\

The GRW model is defined by the following postulates:
\begin{enumerate}  
\item Every physical system is subject to spontaneous localizations (i.e. collapses) in space which take place at random times, following a Poisson distribution with the mean rate given by\footnote{In their original formulation \cite{Ghirardi:1986aa}, Ghirardi, Rimini and Weber considered the possibility that different particles can have different collapse rate $\lambda_i$. However, this is not required and in literature only one $\lambda$, representing the collapse rate for a nucleon, is considered. For composite objects, the corresponding total collapse rate can be calculated through the amplification mechanism discussed below.} $\lambda$.
\item The localization at the point $\bf{a}$ is described as
\begin{equation}
|\psi\rangle\;\;\;\rightarrow\;\;\;\frac{\hat{L}_{\bf a}|\psi\rangle}{||\hat{L}_{\bf a}|\psi\rangle||},
\end{equation}
where the localization operator $\hat{L}_{\bf a}$ is given by 
\begin{equation}
\hat{L}_{\bf a}=(\pi \rC^2)^{-3/4}e^{-\frac{(\hat{\mathbf{q}}-\mathbf{a})^2}{2\rC^2}}.
\end{equation}
\item The probability of having a localization at the point $\bf a$ is $||\hat{L}_{\bf a}|\psi\rangle||^2$. 
\item When there are no localizations in space, the system evolves according the Schr\"{o}dinger equation
\begin{equation}
i\hbar\frac{\D|\psi(t)\rangle}{\D t}=\hat{H} |\psi(t)\rangle.
\end{equation}
\end{enumerate}
We now show how localization works by means of a simple example. Consider a one dimensional system in a superposition of two states which are spatially localized around the points $a$ and $-a$ with $a\gg \rC$. Each state is represented by a wave packet with a width smaller than $\rC$. The total state reads $\Psi(x)=\psi_a(x)+\psi_{-a}(x)$. Let us suppose that a collapse takes place around the point $a$. This amounts in multiplying the wave function by a Gaussian centred in $a$ with width $\rC$ and normalize the resulting state, as dictated by the above postulate 2. Then, after the collapse, the branch of the wave function $\psi_{-a}(x)$ is suppressed and the wave function of the particle is well localized around $a$. This is how, starting from a delocalized wave function, we ended up with a localized one. Note also that the postulate 3 guarantees that the probabilities of having a collapse around the points $a$ or $-a$ is, in a good approximation, 50\%. More in general, the postulate 3 guarantees two fundamental properties: \textit{(i)} in the limit of high number of collapses we get the Born rule, and \textit{(ii)} the master equation associated to the GRW dynamics is linear (see~\cite{Bassi:2003aa} for details), which is a necessary condition to guarantee the not faster-than-light signalling \cite{Gisin:1989aa}.\\

Together with the localizations, there is another fundamental feature required in any collapse model: the amplification mechanism. The amplification mechanism guarantees that, given a composite object, its center of mass localizes with a rate given by the sum of the rates of localization of its elementary constituents. This implies that quantum mechanics is still an excellent approximation for microscopic objects: the collapses are so rare that their effects on the dynamics can be neglected for all practical purposes. Conversely, the effective collapse rate for a macroscopic object is large due to the amplification mechanism, and then any spatial superpositions is rapidly suppressed.\\

{  To understand how the amplification mechanics works, let us consider a rigid system composed by $N$ particles in the following superposition state
\begin{equation}\label{macrosup}
\Psi(x_1,...,x_N)=\psi_a(x_1,...,x_N)+\psi_{-a}(x_1,...,x_N).
\end{equation}
Now, let us suppose that the $j$-th particle collapses around $a$. As in the single particle case, this implies that $\Psi$ gets multiplied by a Gaussian centered in $a$, namely
\begin{equation}
|\Psi\rangle\;\;\;\rightarrow\;\;\;\frac{\hat{L}_{\bf a}^{(j)}|\Psi\rangle}{||\hat{L}_{\bf a}^{(j)}|\Psi\rangle||},
\end{equation}
with $\hat{L}_{\bf a}^{(j)}=(\pi \rC^2)^{-3/4}\exp{-{(\hat{\mathbf{q}}_j-\mathbf{a})^2}/{2\rC^2}}.$
Consequentely, the $\psi_{-a}$ branch of the superposition is suppressed. Since the collapses happen independently for any $j$-th particle, then the state in Eq.~\eqref{macrosup} collapses with an amplified rate $\Lambda= N \lambda$. }

\section{The CSL model}\label{sec:CSL}
In the GRW model, the collapse does not preserve the symmetry of the wave function implying that the model cannot describe identical particles. This limitation was overcome with the CSL model, which was formulated using the second quantization formalism. Thus, it automatically guarantees that its dynamics preserve the wave function symmetry. In this model, the collapse is described by a non-linear interaction with a classical noise. The CSL equation reads \cite{Bassi:2003aa}:
\begin{equation}
\begin{aligned}\label{CSLstoc}
\frac{\D\ket{\psi_t}}{\D t}&=\left[	-\frac i\hbar \hat H+\frac{\sqrt{\lambda}}{m_0}\int \D \x\,\left(\hat M(\x)-\braket{\hat M(\x)}_t\right)w_t(\x)\right.\\
&\left.-\frac\lambda{2m_0^2}\int\D\x\,\left(\hat M(\x)-\braket{\hat M(\x)}_t\right)^2
	\right]\ket{\psi_t},
\end{aligned}
\end{equation}
where $\ket{\psi_t}$ is the $N$ particle wave function and $\hat H$ is the system Hamiltonian. Here $m_0$ is a reference mass taken as that of a nucleon,  and $w_t(\x)$ is the noise providing the collapse, characterized by $
\mathbb E[{w_t(\z)}]=0$ and $\mathbb E[{w_t(\z)w_s(\x)}]=\delta^{(3)}(\z-\x)\delta(t-s)$, where $\mathbb E[\,\cdot\,]$ denotes the stochastic average over the noise.
The locally averaged mass density operator is defined as
\bq\label{def1M}
\hat M(\x)=\sum_j m_j\sum_s\int\D\y\,g(\x-\y)\hat a^\dag_j(\y,s)\hat a_j(\y,s),
\eq
where $\hat a^\dag_j(\y,s)$ and $\hat a_j(\y,s)$ are, respectively, the creation and annihilation operators of a particle of type $j$ with spin $s$ at the point $\y$, while
\bq
g(\x-\y)=\frac{1}{\pi^{3/4}\rC^{3/2}}e^{-\tfrac{(\x-\y)^2}{(2\rC^2)}},
\eq
is a smearing function imposing the spatial correlation of the collapses. 
Exactly as for the GRW model, also in the CSL model the wave function gets localized in space. Indeed, the effect of the second and the third terms in Eq.~\eqref{CSLstoc} is to induce a localization in the eigenstates of the operators $\hat M(\x)$ \cite{Adler:2007aa}, which are position eigenstates. The mass proportionality of $\hat M(\x)$ guarantees automatically the implementation of the amplification mechanism. \\

{Regarding the amplification mechanism, the mass proportionality of $\hat M(\x)$ automatically implements it. However, in CSL model the amplification factor is different compared to that in the GRW model. Indeed, in CSL, the amplification factor depends on the shape of the considered system, and not just on the number $N$ of its nucleons. %For a rigid body, a good formula for estimating this amplification factor was suggested by Adler [] and it is given by $\Lambda=N_s n^2 \lambda$ where $N_s$ represent the number of spheres with radius $r_C$ required to cover the volume of the system and $n$ the number of nucleons inside each sphere. 
In the particular case of a rigid body, when its size is smaller than $\rC$ we have $\Lambda=N^2\lambda$. Conversely, in the limit of $\rC$ smaller that the interparticle distance, the amplification scales with $\Lambda=N\lambda$, which is the same as in the GRW model.}

Working directly with Eq.~\eqref{CSLstoc} is in general problematic, being the equation non-linear. However, as long as we are interested in computing expectation values, we can replace the CSL dynamics with \cite{Adler:2007aa}
\begin{equation}\label{random_uni}
i\hbar\frac{\D|\psi_t\rangle}{\D t}=\left[\hat H-\frac{\hbar\sqrt{\lambda}}{m_0}\int \D\x\, \hat M(\x) w_t(\x)\right]|\psi_t\rangle.
\end{equation}
which is a stochastic Schr\"{o}dinger equation and is much simpler to handle.

\section{Interferometric experiments}\label{int}
\begin{figure*}[b!]
\includegraphics[width=\textwidth]{{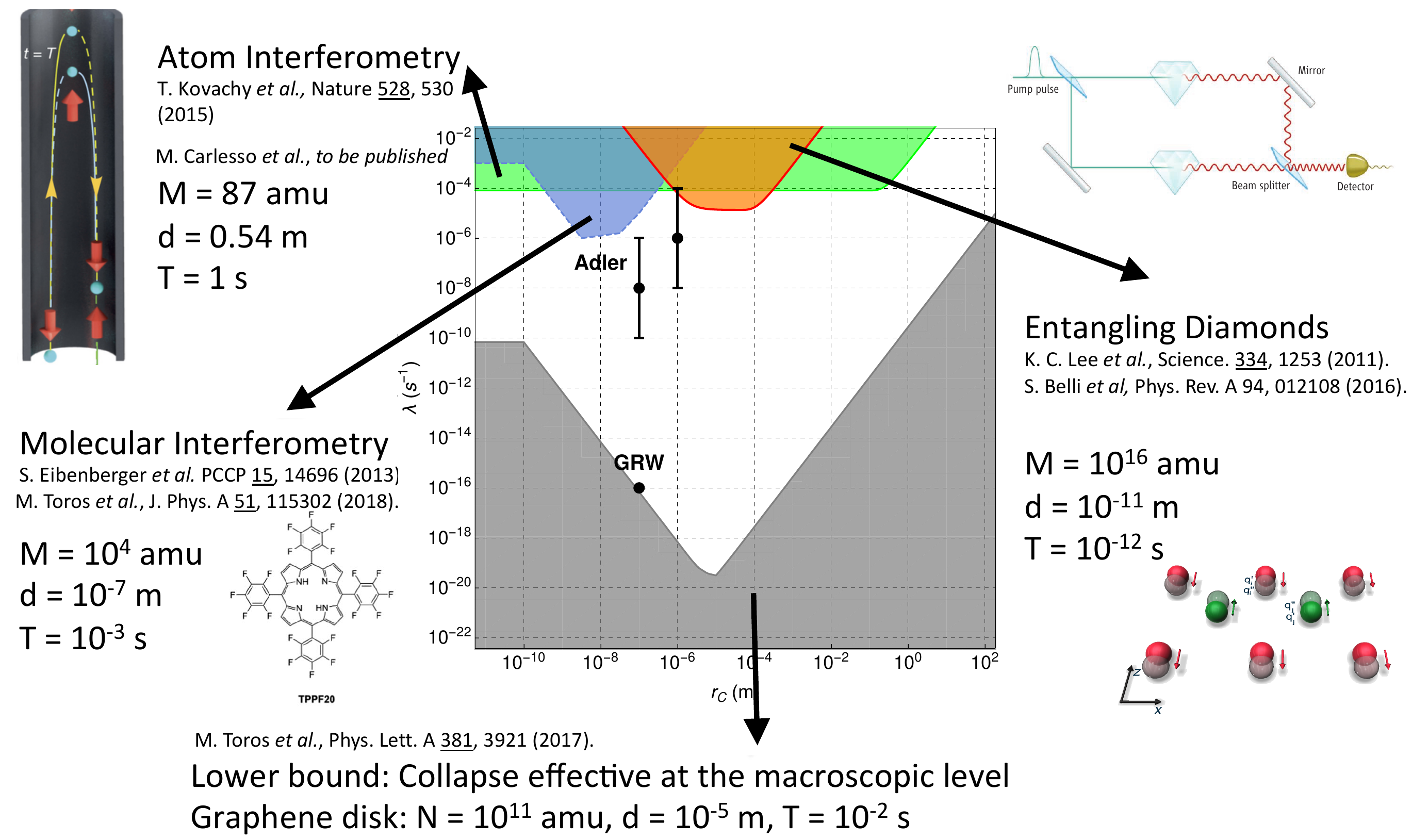}}
\caption{\label{fig:interf} Exclusion plots for the CSL parameters from interferometric experiments with respect to the GRW's and Adler's theoretically proposed values \cite{Ghirardi:1986aa,Adler:2007ab}: molecular interferometry \cite{Eibenberger:2013aa,Toros:2017aa} (blue area), atom interferometry \cite{Kovachy:2015aa} (green area) and experiment with entangled diamonds \cite{Lee:2011aa,Belli:2016aa} (orange area). We report with the grey area the region excluded from theoretical arguments \cite{Toros:2017aa}. $M$, $d$ and $T$ refer respectively to the mass, the superposition distance involved and the time of each experiment.}
\end{figure*}
We can divide the possible tests of collapse models in two classes of experiments: interferometric and non-interferometric ones. Interferometric experiments are the most natural choice of testing collapse models since they detect the direct action of collapse models. One prepares a quantum system in a superposition state and then measures the corresponding interference pattern. The collapse action will be determined by the reduction of the interference contrast. Fig.~\ref{fig:interf} summarizes the state of the art of the bounds on the collapse parameters inferred from interferometric experiments, where different bounds are shown: in green and in blue from cold-atoms \cite{Kovachy:2015aa} and molecular \cite{Eibenberger:2013aa,Hornberger:2004aa,Toros:2017aa,Toros:2018aa} interferometry respectively, and in orange from entanglement experiments with diamonds \cite{Lee:2011aa,Belli:2016aa}. 
{By following the same reasoning, one derives also which is the minimum action that collapse models should impose to actually solve the measurement problem at the macroscopic level. Specifically, a lower bound (grey area) is derived by requiring that a superposition of a single-layered graphene disk of radius $\simeq 10^{-5}$\,m collapses in less than $\simeq 10^{-2}$\,s \cite{Toros:2017aa}.}

\section{Non-interferometric experiments}\label{non_int}

In the second class of possible tests of collapse models, one exploits an indirect effect: the Brownian-like motion induced by the interaction of the collapse noise with the considered system. This motion imposes a growth of the position variance of the center-of-mass of the system, which can be eventually measured. Alternatively, if the system is charged, one can measure the radiation emission due to its acceleration given by such a motion. Since no superposition is involved in these experiments, one can make use of systems of truly macroscopic dimensions. Indeed, due to the in-built amplification mechanism, the collapse effect becomes stronger and thus easier to be detected. 
{However, larger systems are also more affected by environmental noises, which compete with that due to the collapses. Thus, to impose strong bounds on CSL parameters, one seeks for a large mass in an experiment that should be as noiseless as possible.}

\begin{figure*}[th!]
\includegraphics[width=\textwidth]{{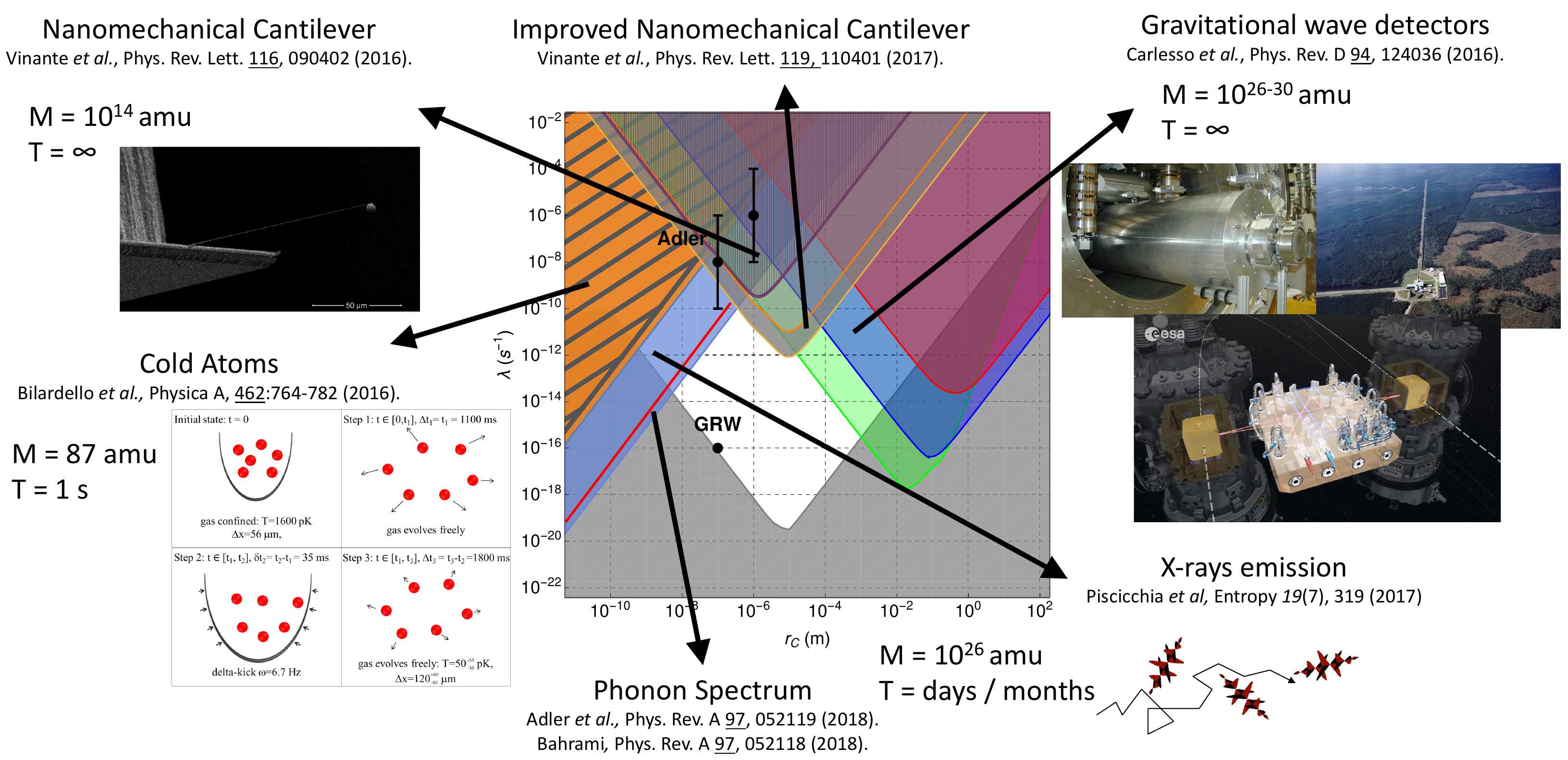}}
\caption{\label{fig:non-interf} Exclusion plots for the CSL parameters from non-interferometric experiments: cold atoms~\cite{Bilardello:2016aa} (orange area), nanomechanical cantilevers \cite{Vinante:2016aa,Vinante:2017aa} (purple and grey-orange bounded areas), gravitational wave detectors AURIGA, LIGO and LISA Pathfinder \cite{Carlesso:2016ac, Helou:2017aa,Carlesso:2018ab} (red, blue and green areas respectively), X-ray measurements~\cite{Aalseth:1999aa,Piscicchia:2017aa} (light blue area), phonon excitations in crystals~\cite{Adler:2018aa,Bahrami:2018aa} (red line). $M$ and $T$ refer respectively to the mass and the time of the experiment.}
\end{figure*}

Fig.~\ref{fig:non-interf} summarizes the state of the art in this class of experiments, which includes experiments involving cold atoms~\cite{Kovachy:2015ab,Bilardello:2016aa}, optomechanical systems~\cite{Usenko:2011aa,Vinante:2016aa,Vinante:2017aa,Vinante:2006aa,Abbott:2016ab,Abbott:2016aa,Armano:2016aa,Carlesso:2016ac, Helou:2017aa,Armano:2018aa}, X-ray measurements~\cite{Aalseth:1999aa,Piscicchia:2017aa}, phonon excitations in crystals~\cite{Adler:2018aa,Bahrami:2018aa}.

Of particular interest is the nanomechanical cantilever experiment described in \cite{Vinante:2017aa}, where an excess noise of known origin was detected. Its value is compatible with that predicted by the CSL model with - up to date - still non-excluded values of the CSL parameters. Several standard mechanisms, able to describe such excess noise, were considered and excluded. An eventual identification of such noise to a standard source will improve the bound of the experiment in \cite{Vinante:2017aa} of one order of magnitude in $\lambda$, see the two orange upper bounds contouring the top grey area in Fig.~\ref{fig:non-interf}.

\section{Generalization of GRW and CSL models}\label{gen}

{There are some limitations on the regime of validity of GRW and CSL models. To make an example, both models are non relativistic. Possible relativistic extensions have been suggested for the GRW model in \cite{Tumulka:2006aa} as well as for the CSL model in \cite{Bedingham:2011aa}.} \\

Moreover, GRW and CSL models have other two weaknesses. The first is the presence of a steady increase in the energy of any system in time, the second is the use of a white (flat) noise.  Here, we discuss how such limitations can be evaded.

\subsection{Dissipative CSL model}
In the CSL model, the energy of any system is not conserved due to the interaction with the noise inducing the collapse. In the case of a free single particle, one has \cite{Bassi:2003aa} 
\begin{equation}\label{en_inc_csl}
\langle \hat H\rangle_{t}=\langle \hat H\rangle_{0}+\frac{3m\lambda\hbar^{2}}{4m_{0}^{2}\rC^{2}}t.
\end{equation}
{The energy of the system grows indefinitely in time. For example, an hydrogen atom is heated by $\simeq 10^{-14}$\,K per year considering the values $\lambda=10^{-16}$\,s$^{-1}$ and $\rC=10^{-7}$\,m.}
Although the increment is small, this feature is not realistic even for a phenomenological model. Here, the CSL noise acts as an infinite temperature bath. Conversely, one expects that a system will eventually thermalize at the finite temperature of the noise. The introduction of dissipation precisely guarantees this.
Indeed, in the dissipative CSL model, Eq.~\eqref{en_inc_csl} becomes
\begin{equation}\label{en_inc_dcsl}
\langle \hat H\rangle_{t}=e^{-\chi t}\left(\langle \hat H\rangle_{0}-H_\text{\tiny as}\right)+H_\text{\tiny as},
\end{equation}
with $\chi=\frac{4km^2\lambda}{(1+k)^5m_0^2}$ and $H_\text{\tiny as}=\frac{3\hbar^2}{16km\rC^2}$, where $k=\frac{\hbar^2}{8m \kb \rC^2T_\text{\tiny CSL}}$. Here, $T_\text{\tiny CSL}$ is a new parameter representing the effective temperature of the noise. Theoretical arguments suggest  $T_\text{\tiny CSL}=1$\,K. \\

For a detailed discussion on the dissipative extension of the CSL (and GRW) model, the reader can refer to \cite{Smirne:2015aa,Smirne:2014aa}. Here, we give an intuition on how dissipation is included in the model. Consider the Fourier transform of the localization operators in the CSL and the dissipative CSL model. They are given respectively by:
\begin{equation}\label{LFT_csl}
\hat M(\mathbf{y})=\sum_{j}\frac{m_{j}}{(2\pi\hbar)^{3}}\int \D\mathbf{P}\D\mathbf{Q}\,e^{-\frac{i}{\hbar}\mathbf{Q}\cdot\mathbf{y}}e^{-\tfrac{\rC^{2}}{2\hbar^{2}}\mathbf{Q}^{2}}\hat a_{j}^{\dagger}(\mathbf{P}+\mathbf{Q})\hat a_{j}(\mathbf{P}),
\end{equation}
and 
\begin{equation}\label{LFT_csl}
\hat M_\text{\tiny D}(\mathbf{y})=\sum_{j}\frac{m_{j}}{(2\pi\hbar)^{3}}\int \D\mathbf{P}\D\mathbf{Q}e^{-\frac{i}{\hbar}\mathbf{Q}\cdot\mathbf{y}}e^{-\tfrac{\rC^{2}}{2\hbar^{2}}|(1+k_{j})\mathbf{Q}+2k_{j}\mathbf{P}|^{2}}\hat a_{j}^{\dagger}(\mathbf{P}+\mathbf{Q})\hat a_{j}(\mathbf{P}).
\end{equation}
Here, the action of the operator $\hat a_{j}^{\dagger}(\mathbf{P}+\mathbf{Q})\hat a_{j}(\mathbf{P})$ is to destroy a particle with momentum $\mathbf{P}$ and to create another one with momentum $\mathbf{P+Q}$, i.e.~to transfer a momentum $\mathbf{Q}$ to the system. In the CSL model, the distribution of the transferred momentum $\mathbf{Q}$ is a Gaussian centered around zero and it does not depend on the system momentum $\mathbf{P}$. This is the reason why the noise keeps heating the system indefinitely. On the contrary, in the dissipative CSL model, the distribution of the possible transferred momentum is centered around a point proportional to $-\mathbf{P}$. In this way, the energy of any system approaches an asymptotic finite value in the longtime regime. \\

Fig.~\ref{fig:extensions} shows the upper bounds to the dissipative CSL extension for different values of $T_\text{\tiny CSL}$. For a more detailed analysis on the current bounds on the dissipative CSL model refer to \cite{Bilardello:2016aa, Toros:2018aa,Toros:2017aa,Nobakht:2018aa}.

\subsection{Colored CSL model}
\begin{figure*}
\centering
\includegraphics[width=0.5\linewidth]{{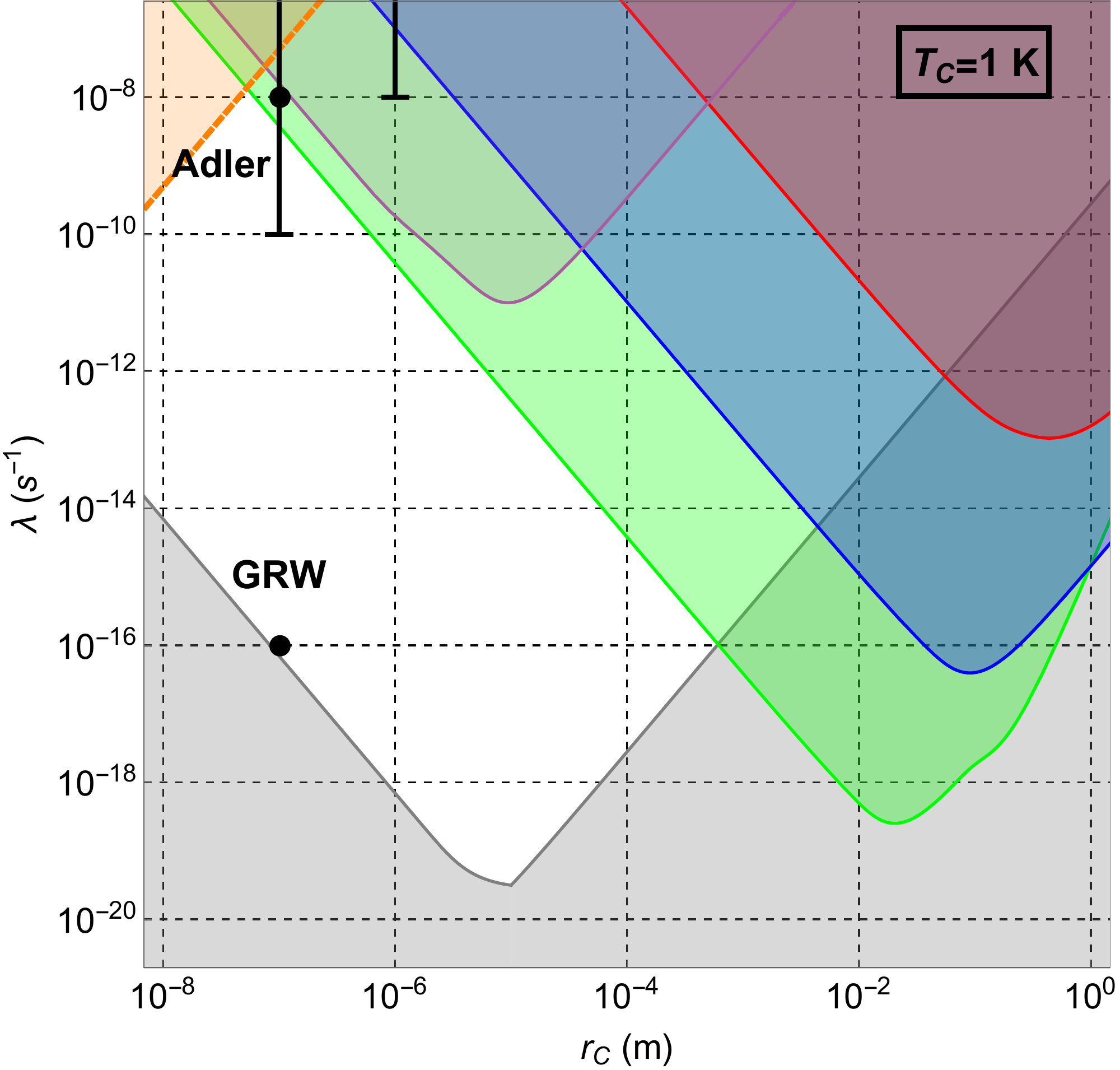}}\includegraphics[width=0.5\linewidth]{{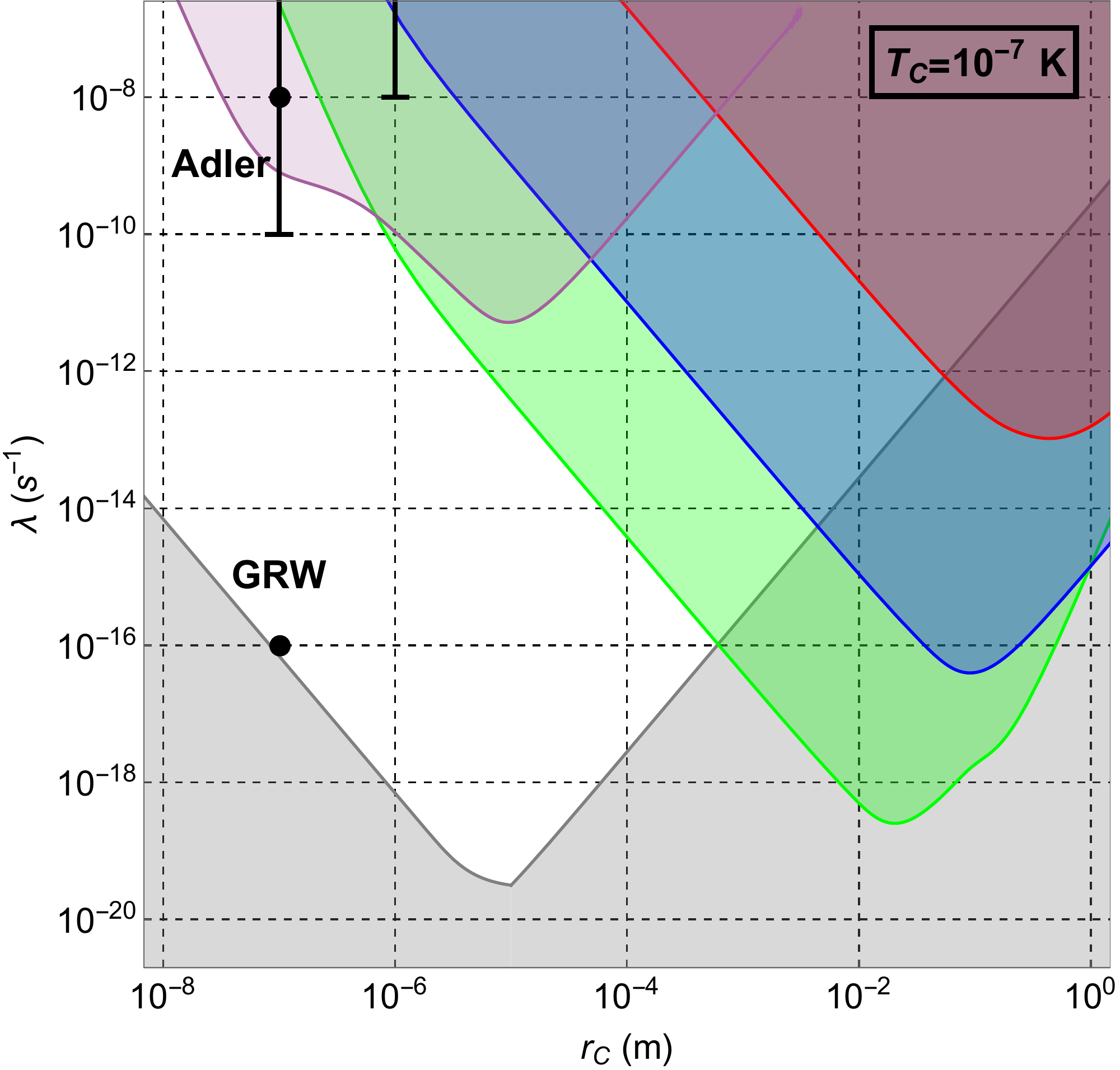}}\\
\includegraphics[width=0.5\linewidth]{{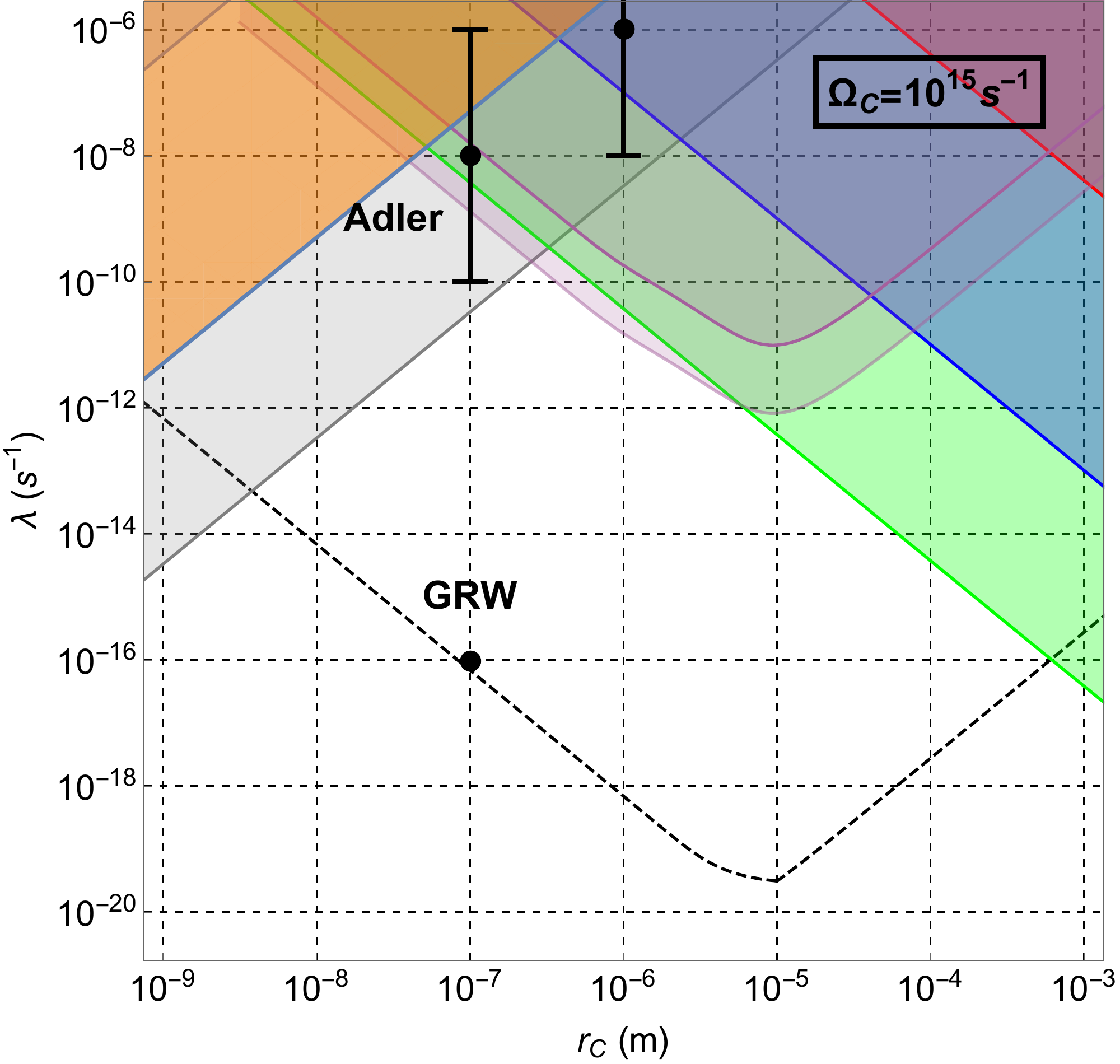}}\includegraphics[width=0.5\linewidth]{{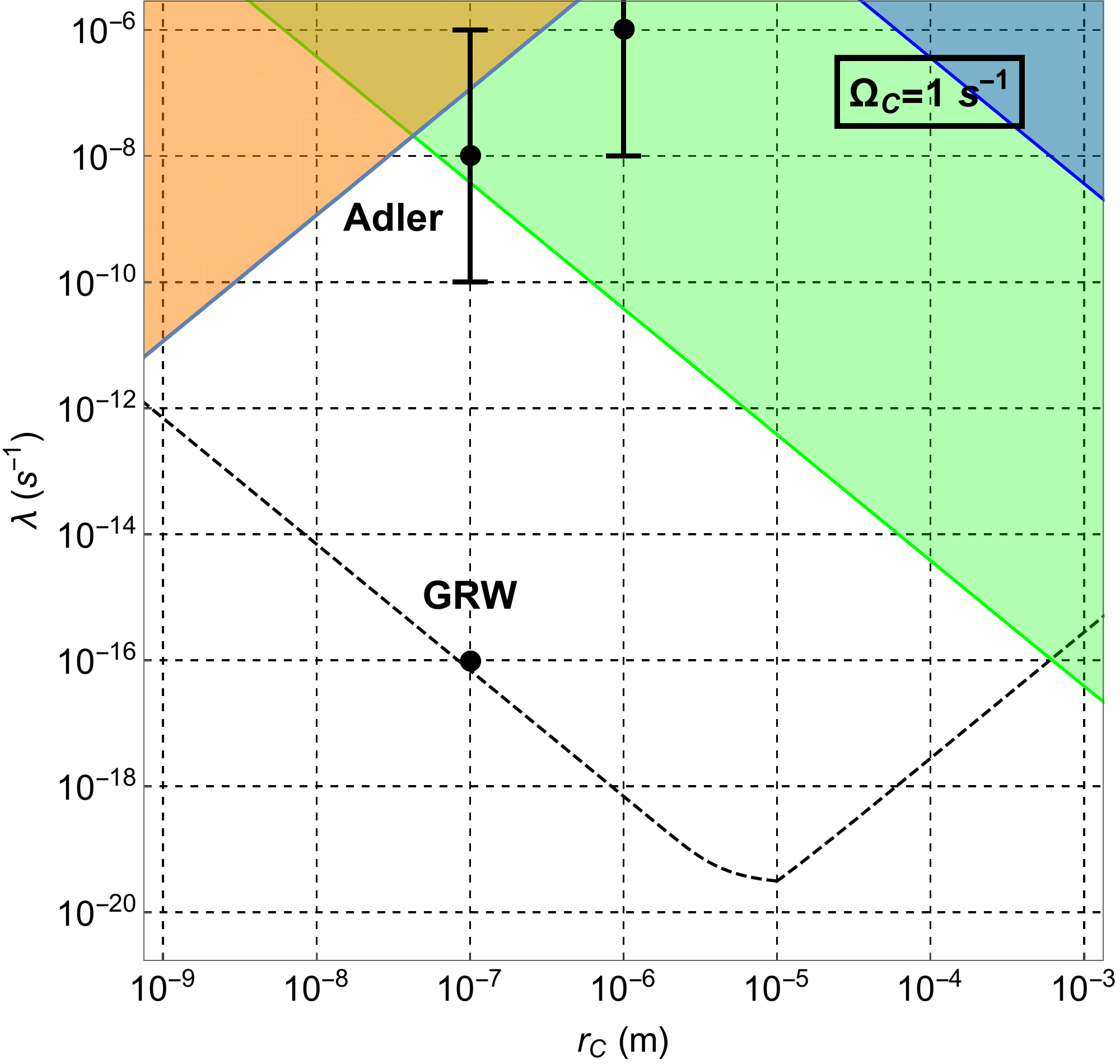}}
\caption{\label{fig:extensions} \textbf{First and second panels}: Upper bounds on the dissipative CSL parameters $\lambda$ and $\rC$ for {two} values of the CSL noise temperature: $T_\text{\tiny CSL}=1\,$K (first panel) and $T_\text{\tiny CSL}=10^{-7}\,$K (second panel). \textbf{Third and fourth panels}: Upper bounds on the colored CSL parameters $\lambda$ and $\rC$ for two values of the frequency cutoff: $\Omega_\text{\tiny c}=10^{15}\,$ Hz (third panel) and $\Omega_\text{\tiny c}=1\,$ Hz (fourth panel). 
Red, blue and green lines (and respective shaded regions): Upper bounds (and exclusion regions) from AURIGA, LIGO and LISA Pathfinder, respectively \cite{Carlesso:2016ac}. Purple region: Upper bound from cantilever experiment \cite{Vinante:2017aa}.  Orange and grey top regions: Upper bound from cold atom experiment \cite{Kovachy:2015ab,Bilardello:2016aa} and  from bulk heating experiments \cite{Adler:2018aa}.
The bottom area shows the excluded region based on theoretical arguments \cite{Toros:2017aa}. }
\end{figure*}
The second limitation of the CSL model is that the noise inducing the collapse is white. While this can be a good approximation in certain regimes, no real noise is expected to be completely white. In particular, it is reasonable that for high enough frequencies the spectrum of the noise presents a cutoff $\Omega_\text{\tiny C}$, whose inverse denotes a characteristic correlation time of the noise. Then, it is important to verify if the presence of a non-white noise affects the model, in particular whether the localization and amplification mechanism are still working. A detailed and analysis for generic collapse equations can be found in \cite{Adler:2007aa,Adler:2008aa}. In general, one can prove that both the aforementioned mechanism work. 
Regarding the predictions of the model, one derives a stochastic Schr\"{o}dinger equation with the same form as Eq.~\eqref{random_uni} where the noise $w_t(\x)$ is substituted by a noise $\xi_t(\x)$ with zero average and correlation $\mathbb E[{\xi_t(\z)\xi_s(\x)}]=\delta^{(3)}(\z-\x)f(t,s)$. Here, $f(t,s)$ denotes the time correlation function. Note that, contrary to the white noise case where the equation is exact, when working with colored noise, Eq.~\eqref{random_uni} is given by a first order expansion in $\lambda$. Since the noise effects are typically small, a perturbative treatment is generally sufficient.\\

{Some experiments are more sensible than others when a colored noise is introduced. For example, the predictions from radiation emission are sensibly modified. Indeed, already a cutoff smaller of the order of $\sim10^{21}$\,Hz suppresses the corresponding bound \cite{Adler:2013aa,Donadi:2014aa,Bassi:2014aa}} Bounds on the CSL parameters for colored noise were studied in detail in \cite{Bilardello:2016aa,Toros:2017aa, Carlesso:2018aa}. In particular, one finds out that the upper bounds from experiments at high frequencies (or involving small time scales) are weakened more and more when moving to smaller value of $\Omega_\text{\tiny C}$. Theoretical arguments suggest $\Omega_\text{\tiny C}\sim10^{12}\,$Hz. Fig.~\ref{fig:extensions} shows the upper bounds to the colored CSL extension for different values of $\Omega_\text{\tiny C}$.

\section{Proposals for future testing}\label{prop}

To confirm or falsify the possibility that the excess noise measured in \cite{Vinante:2017aa} is actually the effect of a collapse mechanism, one needs to consider new experimental techniques for an independent inquiry.

\begin{figure*}
\centering
\includegraphics[width=0.5\linewidth]{{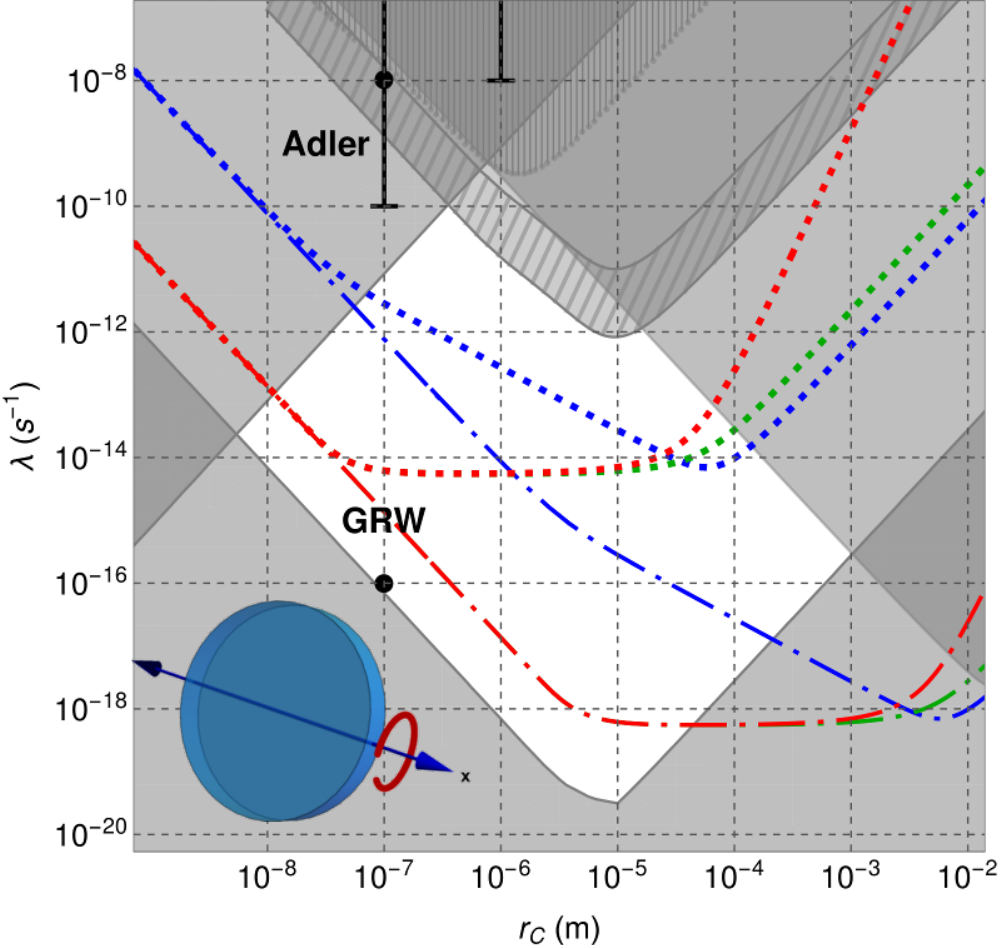}}\includegraphics[width=0.5\linewidth]{{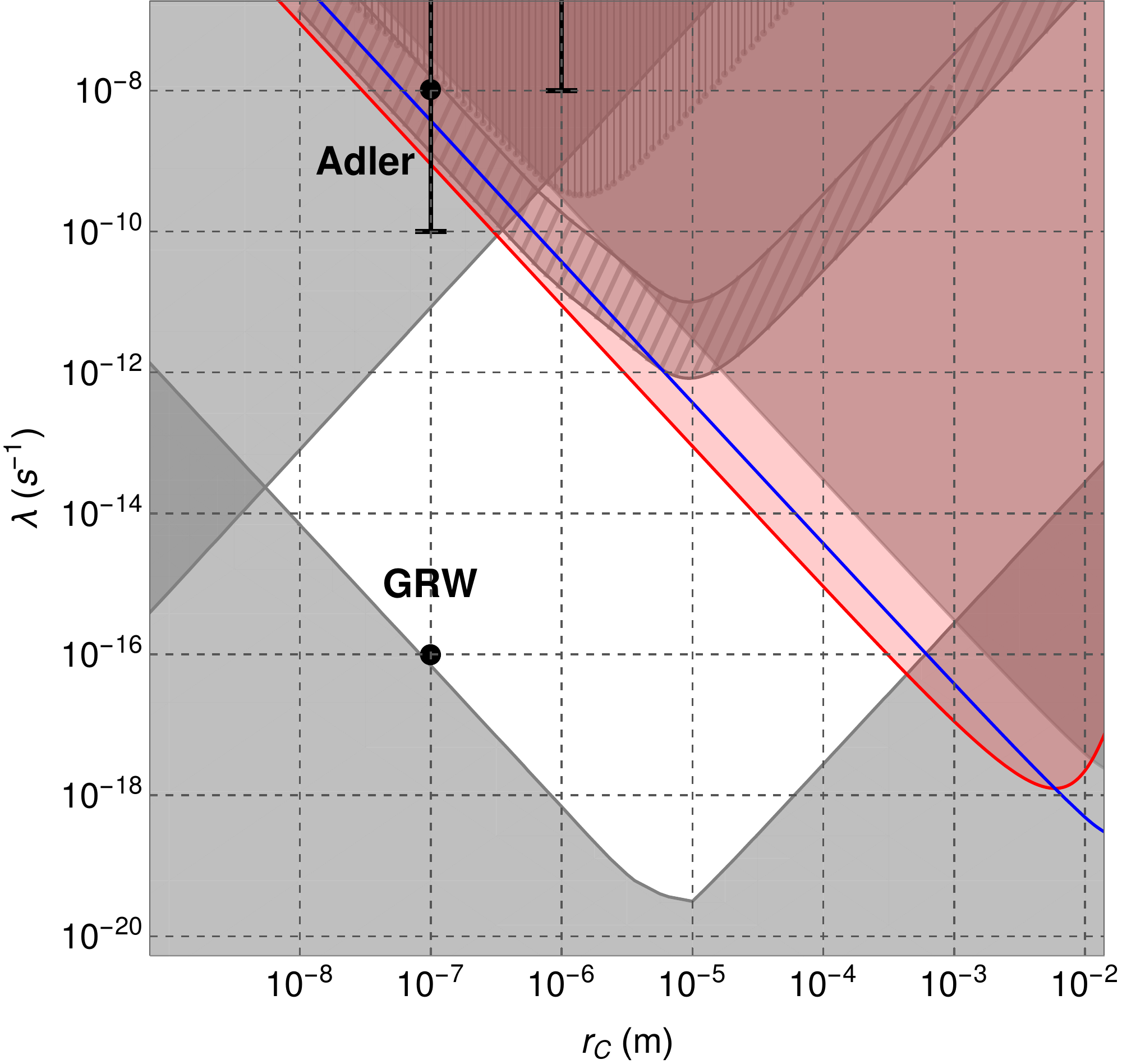}}
\includegraphics[width=0.5\linewidth]{{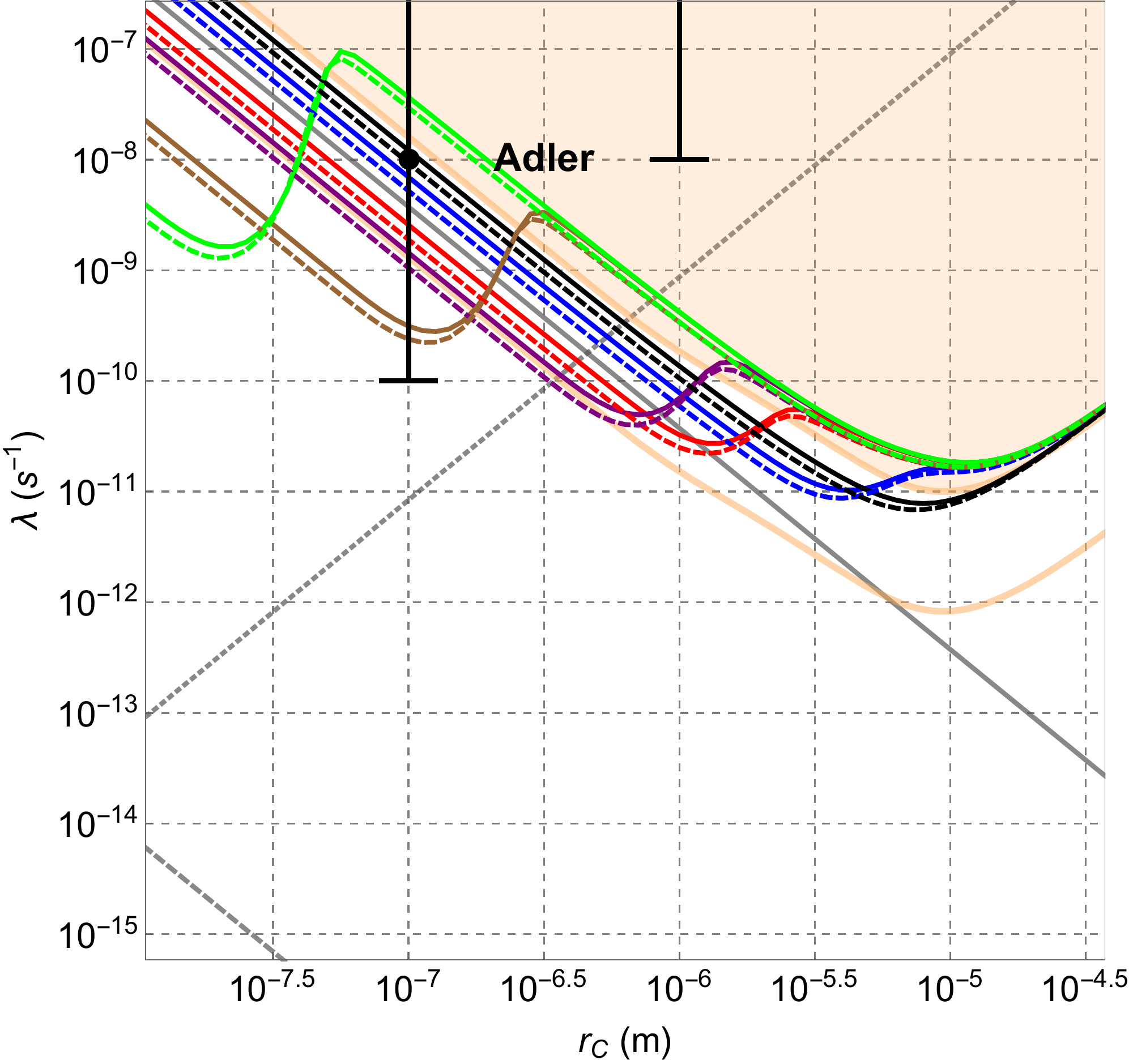}}\includegraphics[width=0.5\linewidth]{{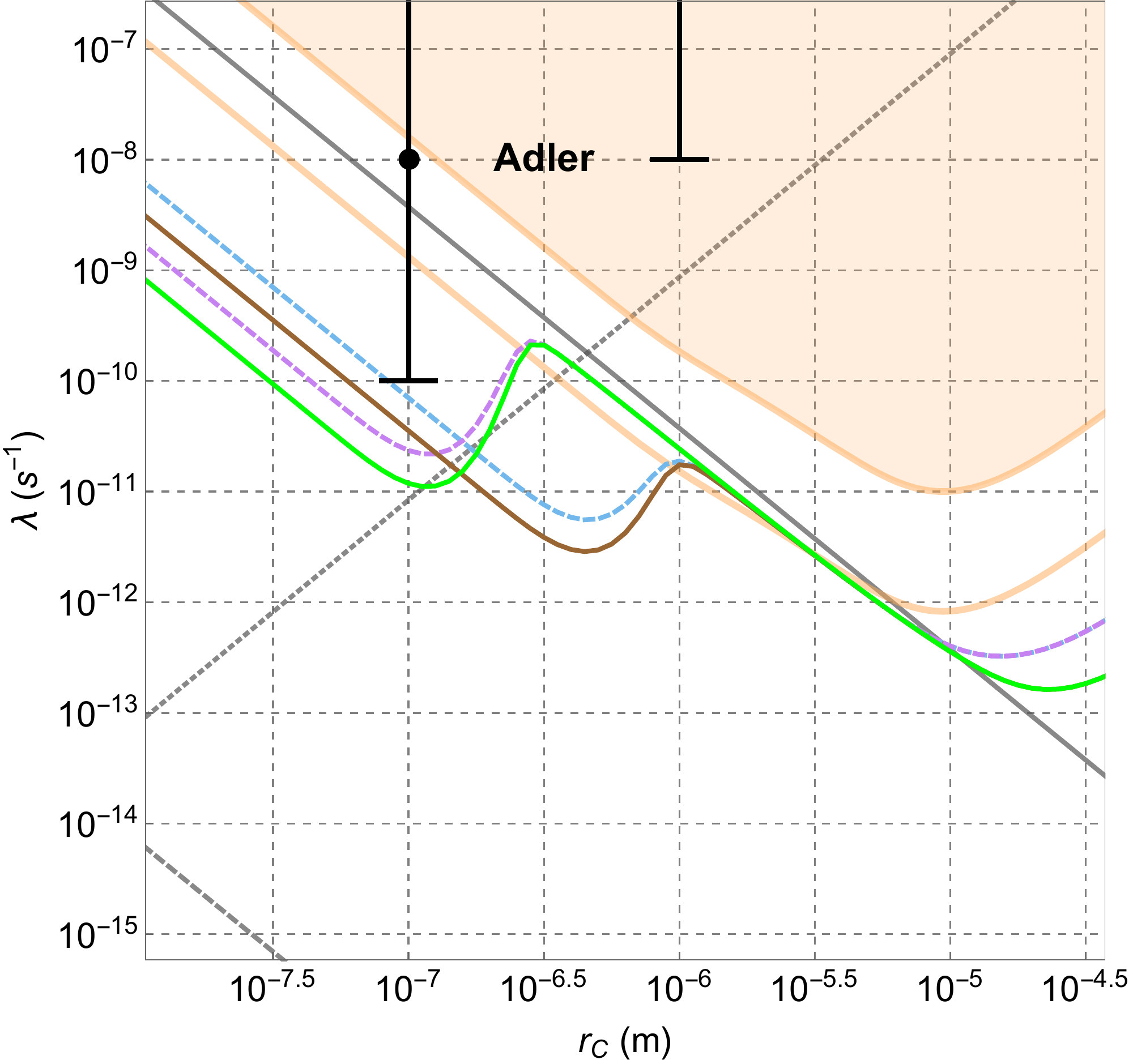}}
\caption{\label{fig:proposals} Exemplification of two possible experimental tests of collapse models. \textbf{First panel}: Results of the analysis proposed in \cite{Schrinski:2017aa,Carlesso:2018ab} where the rotational degrees of freedom of a cylinder are studied. The red line denotes the upper bound that can be obtained from the constrains given by the rotational motion, compared with those from the translations (blue and green lines). \textbf{Second panel}: Red shaded area highlights the hypothetical excluded value of the collapse parameters that could be to derived from the conversion of the translational noise of LISA Pathfinder to rotational one \cite{Carlesso:2018ab}. This is compared to the new (old) upper bounds from the translational motion shown with the blue line \cite{Carlesso:2018ab} (grey area \cite{Carlesso:2016ac}). \textbf{Third panels}: Hypothetical upper bounds obtained from substituting the sphere attached to the cantilever used in \cite{Vinante:2017aa} with a multilayer cuboid of the same mass for various thickness of the layers \cite{Carlesso:2018ac}. The bounds are compared with that from the improved cantilever experiment \cite{Vinante:2017aa} shown in orange. \textbf{Fourth panel}: Same as the third panel, but with a mass ten times larger.}
\end{figure*}

One possible test consists of focusing on the rotational degrees of freedom of a system and its collapse-induced Brownian motion \cite{Schrinski:2017aa,Carlesso:2018ab}. It turns out that for truly macroscopic systems, this technique can provide a sensible improvement of the bounds on the collapse parameters, cf.~Fig.~\ref{fig:proposals}. A direct application was considered in \cite{Carlesso:2018ab}, where the bound from LISA Pathfinder \cite{Carlesso:2016ac} can be significantly improved by considering also the rotational degrees of freedom.

Another proposal \cite{Carlesso:2018ac} considered a modification of the cantilever experiment in \cite{Vinante:2017aa} where the homogeneous mass is substituted with one made of several layers of two different materials. This will increment the CSL noise for the values of $\rC$ of the order of the thickness of the layers. An example is shown in Fig.~\ref{fig:proposals}.

These are just two of the several proposals \cite{Collett:2003aa,Goldwater:2016aa,Kaltenbaek:2016aa,McMillen:2017aa,Mishra:2018aa} suggested over the last years to push the exploration of the CSL parameter space.

\section{Conclusions}

{We discussed how collapse models provide a solution to the measurement problem. They modify the Schr\"{o}dinger dynamics introducing a spatial collapse of the wave function. We focused in particular on the most relevant collapse models, which are the GRW and the CSL model. We discussed their main properties and the status of the experimental bounds on their phenomenological parameters $\lambda$ and $\rC$ (Fig.~\ref{fig:interf} and Fig.~\ref{fig:non-interf} ). In particular, non-interferometric experiments provide the strongest tests of collapse models. They extend over a broad set of possible systems, which differ in size, form, materials, degrees of freedom and much more. Moreover, we considered the dissipative and colored noise extensions of the CSL model. Also in these cases, non-interferometric tests are the most relevant for the experimental investigation (Fig.~\ref{fig:extensions}). Finally, we present several non-interferometric proposals that were suggested to push further the exploration of collapse models (Fig.~\ref{fig:proposals}).}

\subsection*{Acknowledgments}
{MC acknowledges the financial support from the H2020 FET Project TEQ (grant n.~766900) and the support from the COST Action QTSpace (CA15220), INFN and the University of Trieste. SD acknowledges the financial support from the Fetzer Franklin Fundation and the support from the COST Action QTSpace (CA15220) and the Frankfurt Institute for Advanced Studies (FIAS).
Both the authors are grateful for the support offered by the WE-Heraeus-Stiftung for the WE-Heraeus-Seminars entitled \textit{``Advances in open systems and fundamental tests of quantum mechanics''}.}

\end{document}